\documentclass[10pt]{article}

\usepackage{amsmath}

\usepackage{array}

\usepackage{appendix}

\usepackage{tocloft}                   % Adds Part structure to table of contents

\usepackage{graphicx}

\usepackage{amsfonts}

\usepackage{amssymb}

\usepackage{mathrsfs}

\usepackage{yfonts}

\usepackage{euscript}

\usepackage{centernot}                 % Permits me to trivially make a centred not version of any object

\usepackage{ifsym}                     % Gives a nicer sharp

\usepackage{lmodern}                   % Gives bold typewriter fonts

\usepackage{upgreek}

\usepackage{mathtools}

\usepackage{color}

\usepackage{slantsc}
\usepackage{calligra}

\usepackage{bbold}          % Gives mathbb acting on lower case and, hopefully, numbers...

\usepackage[T1]{fontenc}

\usepackage{epsf}

\usepackage{latexsym}

\usepackage{tipa}

\usepackage{makeidx}

\makeindex

\def\b{\begin{equation}}

\def\e{\begin{equation}}

\def\be{\begin{equation}}              % Longer older ones kept for rext import compatibility.

\def\ee{\end{equation}}

\def\beq{\begin{equation}}

\def\eeq{\end{equation}}

\def\bea{\begin{eqnarray}}

\def\eea{\end{eqnarray}}

\def\m{\mbox{ }}

\def\mma {\m , \m \m }

\def\!{\hspace{-1.6667em}}

\def\n{\noindent}

\def\f{\footnote}

\def\u{\underline}

\def\w{\widetilde}

\def\slLambda{\mathit{\Lambda}}                   % Cosmological Constant 

\def\slOmega{\mathit{\Omega}}                      % Misner variable
\def\uix{\u{x}}

\def\bis{\mbox{\boldmath$s$}}

\def\biC{\mbox{\boldmath$C$}}              % Index-free form of the Poisson tensor

\def\biE{\mbox{\boldmath$E$}}

\def\biP{\mbox{\boldmath$P$}}

\def\biQ{\mbox{\boldmath$Q$}}

\def\sbiU{\mbox{\ttfamily\fontseries{b}\selectfont U}}                   %

\def\sbiB{\mbox{\ttfamily\fontseries{b}\selectfont B}}

\def\sbiD{\mbox{\ttfamily\fontseries{b}\selectfont D}} 

\def\sbiM{\mbox{\scriptsize\boldmath$M$}}

\def\mD{\mbox{D}}                        % Double collision/ points in triangleland, can be higher-d regions in other models.

\def\mM{\mbox{M}}                        % Mergers

\def\mN{\mbox{N}}

\def\mh{\mbox{h}}

\def\mp{\mbox{p}}

\def\bh{\u{\u{\mbox{h}}}  }            %  The 3-metric spatial-index-less.  

\def\urho{{\u{\rho}}}

\def\bK{\mbox{\bf K}}

\def\bN{\mbox{\bf N}}

\def\bQ{\mbox{\bf Q}}

\def\bg{\mbox{\bf g}}

\def\bh{\mbox{\bf h}}

\def\bp{\mbox{\bf p}}

\def\bupSigma{\mbox{\boldmath$\Sigma$}}                 % Spatial hypersurface topological manifold
\def\sbupSigma{\mbox{\scriptsize\boldmath$\Sigma$}}     % subscript version of the previous

\def\scbL{\mbox{\boldmath\scriptsize ${\cal L}$}}

\def\fC{\mbox{\sffamily C}}

\def\sa{\mbox{\scriptsize a}}

\def\scc{\mbox{\scriptsize c}}

\def\se{\mbox{\scriptsize e}}

\def\sf{\mbox{\scriptsize f}}

\def\si{\mbox{\scriptsize i}}

\def\sll{\mbox{\scriptsize l}}  % NB EXCEPTIONAL DEF as \sl is reserved for slant.

\def\sm{\mbox{\scriptsize m}}

\def\sn{\mbox{\scriptsize n}}

\def\sss{\mbox{\scriptsize s}}  %TO AVOID ARXIV changing \ss to German double s.

\def\su{\mbox{\scriptsize u}}

\def\sv{\mbox{\scriptsize v}}

\def\sE{\mbox{\scriptsize E}}

\def\sG{\mbox{\scriptsize G}}

\def\sH{\mbox{\scriptsize H}}

\def\sJ{\mbox{\scriptsize J}}

\def\sM{\mbox{\scriptsize M}}

\def\sQ{\mbox{\scriptsize Q}} 

\def\sR{\mbox{\scriptsize R}}

\def\sU{\mbox{\scriptsize U}}

\def\sfS{\mbox{\sffamily{\scriptsize S}}}      % index for secondary constraints

\def\sbM{\mbox{{\bf \scriptsize M}}}

\def\sbN{\mbox{{\bf \scriptsize N}}}

\def\sbcC{\mbox{\boldmath \scriptsize ${\cal C}$}}

\def\sbcF{\mbox{\boldmath \scriptsize ${\cal F}$}}

\def\sbcG{\mbox{\boldmath \scriptsize ${\cal G}$}}

\def\sbcL{\mbox{\boldmath \scriptsize ${\cal L}$}}

\def\sbcP{\mbox{\boldmath \scriptsize ${\cal P}$}}

\def\sbcM{\mbox{\boldmath \scriptsize ${\cal M}$}}

\def\tf{\mbox{\tiny f}}

\def\ti{\mbox{\tiny i}}

\def\tn{\mbox{\tiny n}}

\def\bscC{\mbox{{\boldmath \scriptsize${\cal C}$}}}                               % Class constraints

\def\bscF{\mbox{{\boldmath \scriptsize${\cal F}$}}}                               % First-class constraints

\def\cr{\mbox{\scriptsize{\bf $\m  \times \m $}}}

\def\sumi2{\sum\mbox{}_{\mbox{}_{\mbox{\scriptsize $i$=1}}}^2}

\def\sumi3{\sum\mbox{}_{\mbox{}_{\mbox{\scriptsize $i$=1}}}^3}

\def\sumAn{\sum\mbox{}_{\mbox{}_{\mbox{\scriptsize $A$=1}}}^{n}}

\def\sumABcycles3{\sum\mbox{}_{\mbox{}_{\mbox{\scriptsize cycles $A,B$=1}}}^{3}}

\def\sumCDcycles3{\sum\mbox{}_{\mbox{}_{\mbox{\scriptsize cycles $C,D$=1}}}^{3}}

\def\sumIN{\sum\mbox{}_{\mbox{}_{\mbox{\scriptsize $I$=1}}}^{N}}

\def\sumj3{\sum\mbox{}_{\mbox{}_{\mbox{\scriptsize $j$=1}}}^3}

\def\sumk3{\sum\mbox{}_{\mbox{}_{\mbox{\scriptsize $k$=1}}}^3}

\def\prodiA1{\prod\mbox{}_{\mbox{}_{\mbox{\scriptsize $i$=1}}}^{A - 1}}

\def\d{\textrm{d}}                                                  % ordinary derivative

\def\pa{\partial}                                                   % partial derivative

\def\es{\m = \m}

\def\peq{\m \mbox{`='} \m}

\def\:={\m := \m}

\def\=:{\m =: \m}

\def\FrT{\mathfrak{T}}                                         % A time-line. 

\def\FrU{\mbox{$\mathfrak{U}$}}                                % Open set, with $\{\FrU_{\sfC}\}$ then being an open cover.  
\def\bFrF{\mbox{\boldmath$\mathfrak{F}$}}                            % 

\def\Frm{\mbox{\Large $\mathfrak{m}$}}                         % The spacetime manifold
\def\lFrg{\mbox{\Large$\mathfrak{g}$}}                         % Irrelevant group, Lie group.
\def\FrT{\mbox{\boldmath$\mathfrak{T}$}}                       % Used for tangent space and then cotangent space is $^*$ of this. 
\def\Hilb{\mbox{{\boldmath$\mathfrak{H}$}ilb}}                 % Hilbert space

\def\scC{\mbox{\scriptsize ${\cal C}$}}                    % general constraint, regardless of linearity or not in the momenta
\def\scE{\mbox{\scriptsize ${\cal E}$}}                    % mechanical energy constraint

\def\scH{\mbox{\scriptsize ${\cal H}$}}                    % Hamiltonian constraint of GR.

\def\scM{\mbox{\scriptsize ${\cal M}$}}                    % momentum constraint of GR  

\def\scP{\mbox{\scriptsize ${\cal P}$}}

\def\scQ{\mbox{\scriptsize ${\cal Q}$}}                    % a second supersymmetric constraint, in theories possessing a such. 

\def\bFlin{\sbcF\mbox{\bf lin}} 

\def\Quad{\scQ\mbox{uad}}                                  % quadratic constraint.

\def\Chronos{\scC\mbox{hronos}}                            % Chronos constraint.

\def\bGauge{\sbcG\mbox{\bf auge}} 

\def\beables{\mbox{\ttfamily\fontseries{b}\selectfont B}} 

\def\chronos{\mbox{\ttfamily\fontseries{b}\selectfont C}} 
					   
\def\Dirac{\mbox{\ttfamily\fontseries{b}\selectfont D}}    % the index-free presentation thereof.  
														   
\def\gauge{\mbox{\ttfamily\fontseries{b}\selectfont G}}                  % the index-free presentation thereof.  

\def\Kuchar{\mbox{\ttfamily\fontseries{b}\selectfont K}}                  % the index-free presentation thereof.  
\def\FrQ{\mbox{\Large $\mathfrak{q}$}}                               % Configuration space
\def\Phase{\mbox{{\boldmath$\mathfrak{P}$}hase}}                     % Phase space.

\def\bFrR{\mbox{\boldmath$\mathfrak{R}$}}                            % First letter of RigPhase, also used for Riem etc.  Is also, by itself, a ring.
\def\Rig-Phase{\bFrR\mbox{ig-}\Phase}                                % Rigged Phase Space
                                                                													   
\def\bFrR{\mbox{\boldmath$\mathfrak{R}$}}                            % Used in regularity structure symbol

\def\bFrR{\mbox{\boldmath$\mathfrak{R}$}}                            % Used in incipient regularity structure symbol

\def\1mat{\u{\u{1}}}                                                 % unit-entry matrix

\def\Mini{\mbox{{\boldmath$\mathfrak{M}$}ini}}                       % Minisuperspace
\def\Positive-Modespace{\mbox{{\boldmath$\mathfrak{M}$}odespace$^+$}}% Positive modespace

\def\POSITIVE-MODESPACE{\mbox{{\boldmath$\mathfrak{M}$}ODESPACE$^+$}}% Positive modespace alongside scalar field matter inhomogeneous modes.
                                                                                                                             														
\def\Riem{\bFrR\mbox{iem}}                                           % Riem

\def\FrO{\mbox{$\mathfrak{O}$}}                                      % Individual gauge orbits.

\def\Kin-Hilb{\mbox{{\boldmath$\mathfrak{K}$}in-\Hilb}}                     % Dynamical Hilbert space 

\def\Mid-Hilb{\mbox{{\boldmath$\mathfrak{M}$}id-\Hilb}}                     % Dynamical Hilbert space 

\def\Dyn-Hilb{\mbox{{\boldmath$\mathfrak{D}$}yn-\Hilb}}                     % Dynamical Hilbert space 

\def\5Star{\mbox{\Large$\star$}}              % Rectified time derviative actually used

\def\K{Kucha\v{r} }

\def\Frr{\mbox{$\mathfrak{r}$}}

\begin{document}

\begin{center}

\Huge{\bf A LOCAL RESOLUTION OF}

\Huge{\bf THE PROBLEM OF TIME}

\Large{\bf IV. Quantum outline and piecemeal Conclusion} 

{\large \bf E.  Anderson}$^1$ 

{\large \it based on calculations done at Peterhouse, Cambridge} 

\end{center}

\begin{abstract}

In this final piecemeal treatment of local Problem of Time facets, and underlying Background Independence aspects, 
we first reconsider the ten local facets and aspects considered so far at the quantum level.
This is essential both to appreciate past conceptualization and naming for these aspects and facets, 
and because the quantum Problem of Time is a more advanced goal than its classical counterpart.

The case is made that each piecemeal aspect of Background Independence can be incorporated, 
i.e.\ each piecemeal facet of the Problem of Time is resolvable, in a local classical sense, using just Lie's Mathematics. 
This case is then extended in Articles V to XIII to a joint resolution, 
i.e.\ successfully handling the lion's share of the Problem of Time that resides in facet interferences.  
Relationalism, Closure, Observables and Constructions are argued to be super-aspects ordered to form a claw alias 3-star 
digraph with Closure as nexus. 
This is realized twice: faithfully for spacetime and with some modification in the split-Relationalism canonical case.  
There is finally a Wheelerian 2-way route between the two realizations, with Spacetime Construction from Space in one direction 
                                                              and Refoliation Invariance in the other. 

\end{abstract}

$^1$ dr.e.anderson.maths.physics *at* protonmail.com

%====================================================================================================================================================================================                       
%====================================================================================================================================================================================                       
\section{Introduction}\label{QBI-Underpin} 
%====================================================================================================================================================================================
%====================================================================================================================================================================================                       

We next consider the Problem of Time \cite{Battelle, DeWitt67, Dirac, Dirac49, K81, K91, K92, I93, K99, APoT, FileR, APoT2, AObs, AObs3, APoT3, ALett, ABook, A-CBI, I, II, III} 
and underlying Background Independence in the more involved quantum-level setting.  
For now, we outline the quantum counterparts of the preceding three Articles' piecemeal consideration of facets and aspects.
This follows on in particular from work of Dirac, Wheeler, DeWitt, \K, and Isham, 
with some further reconceptualizations by the Author.  

\m 

\n Passage to Quantum Theory usually departs from Newtonian Mechanics or SR.  
Attempting to extend this to GR amounts to a `Background Dependence versus Independence' Paradigm Split, 
in which GR and Ordinary Quantum Theory lie on opposite sides.
\n Many of the more difficult parts of the Problem of Time occur because the `time' of Background Independence GR and
                                                                         the `time' of the ordinary Background Dependent Quantum Theory are mutually incompatible notions.
This causes difficulties in trying to replace these two branches of Physics with a single framework in regimes in which neither Quantum Theory nor GR can be neglected. 
 
\m 

\n This situation arose historically by each of these two areas of Physics developing in a different direction both conceptually and technically, 
without enough cross-checks to keep Physics within a single overarching Paradigm. 
This Paradigm Split has a further practical justification which continues to apply today: 
that our practical experiences are of regimes that involve at most one of QM or GR.
Indeed, regimes requiring both of these at once would feature the decidedly outlandish Planck units \cite{ABook} of Quantum Gravity 
\cite{DeWitt67, I74, I81, KieferBook, ABook}, as would be required for parts of Black Hole Physics and Early-Universe Cosmology.

\m 

\n GR can be viewed as not only a Relativistic Theory of Gravitation 
but also as a freeing from absolute or Background Dependent structures \cite{Dirac, K92, I93, PrimaFacie, I95, BI00b, Carlip01, RovelliBook, KieferBook}.  
Indeed, from this gestalt perspective of GR, the wording `Quantum Gravity' is itself is a misnomer 
since it refers solely to GR in its aspect as a Relativistic Theory of Gravitation.  
Quantum Background Independence \cite{ABook} has a comparable status, 
and only specifically Background-Independent Quantum Gravity carries over this gestalt status to the quantum domain.  

\m

\n The Problem of Time is moreover pervasive throughout sufficiently GR-like attempts at formulating Quantum Gravity, at both the quantum and classical levels. 
For now, we take the geometrodynamical and spacetime formulations of GR to be representative, and concentrate on these.
See Parts II and III of \cite{ABook} 
for commentary on the Problem of Time and Background Independence in other Quantum Gravity programs \cite{Ashtekar, ThiemannBook, Sugra, Polchinski}.  

\m 

\n Sec 2 sketches Kinematical Quantization \cite{Mackey, I84}; as a preliminary Assignment of Observables, 
this is our main exception to aspects being approached in the same order as at the classical level.  
Sec 3 covers quantum-level Temporal Relationalism, the corresponding Problem of Time facet for which is the well-known Frozen Formalism Problem \cite{K92, I93}. 
Sec 4 entertains quantum-level Configurational Relationalism. 
Sec 5 considers quantum-level Constraint Closure, 
whose associated Problem of Time facet Isham and \K termed Functional Evolution Problem in the field-theoretic context. 
Sec 6 contemplates quantum-level Assignment of Observables, now taking into consideration constraints as well. 
Sec 7 outlines Quantum Spacetime Construction from Space, 
after which Sec 8 and 9 indulge quantum-level Spacetime Relationalism (including in Path Integral Approaches), Generator Closure and Assignment of Observables. 
Sec 10 touches upon Refoliation Invariance. 

\m 

\n Sec 11 is the piecemeal facets Conclusion (to Articles I to IV's account, thus including consideration of this Series' main classical case as well).    
This maps out this series' progression 
from Kucha\v{r} and Isham's Problem of Time facets to the Author's Background Independence aspects, 
and gives classical- and quantum-level orders for these to be approached in.
Finally, it emphasizes the centrality of Lie's Mathematics \cite{Lie, Serre-Lie, Lee2} to the whole of this venture, 
while deducing the cental role played by Closure in all of Lie Theory, A Local Resolution of the Problem of Time, and a Local Theory of Background Independence.   
This is already apparent from Closure being the nexus -- connecting middle point -- of the Lie star digraph, 
two copies of which (canonical and spacetime), plus a Wheelerian 2-way route therebetween, resolve the Problem of Time's facet-ordering problem \cite{K92, I93, K93}.

%====================================================================================================================================================================================
%=========================================================================================================================================================================================
\section{Kinematical Quantization (aspect 4)}\label{Kin-Quant} 
%====================================================================================================================================================================================
%====================================================================================================================================================================================

In Quantum Theory, observables carry the further connotation of being self-adjoint operators $\widehat{\beables}$. 
By this, their eigenvalues are real-valued and so can correspond to measured or experienced physical properties.

\m 

\n{\bf Structure 1} Kinematical quantization \cite{Mackey, I84} comes first, so as to provide us with incipient operators. 
This is already nontrivial to pick as most phase space functions 
\be 
F(\biQ, \biP)
\ee 
cannot be consistently promoted to quantum operators
In this way, we identify this as finding unrestricted observables
\be 
\widehat{\sbiU}  \m 
\ee 
forming the space
\be 
\FrU\mbox{-}\FrO\mp
\ee 
{\bf Example 1} This is the role $\widehat{p}$ and $\widehat{q}$ play in 1-$d$, 
or $\widehat{\u{p}}$, $\widehat{\u{q}}$ and angular momentum $\widehat{\u{J}}$ play in 3-$d$, in each case for single-particle QM.  

\m 

\n{\bf Remark 1} The point of kinematical quantization is what operators can consistently assume this mantle for QM on other topologies and geometries 
than flat space.  
Kinematical quantum operators act on an incipient kinematical Hilbert space of wavefuntions. 

\m 

\n{\bf Remark 2} This is but an incipient space of operators since constraints are not yet taken into account. 
One cannot proceed without this in the presence of constraints, since one needs the kinematic quantization operators to build the quantum constraints out of. 
Once the constraints' wave equations are taken into account, one can supplant the incipient kinematical Hilbert space with the actual physical Hilbert space.  

\m 

\n{\bf Remark 3} All other Background Independence and Problem of Time matters follow moreover in the classical version's ordering of aspects-or-facets, 
which thus conceptually and technically pre-empts the quantum version.  

\m 

\n{\bf Remark 4} Outlining specific examples of this require consideration of Interference with Configurational Relationalism: reduced versus Dirac quantization. 
This is because if classical reduction of $\bFlin$ has already been carried out, kinematical quantization is distinct from if it has not.  

\m 

\n{\bf Aside 1} Dealing with $\Chronos$ classically as well would be an alternative tempus ante quantum treatment, 
which becomes theory-dependent rather than universal and is not treated here.  

\m 

\n{\bf Example 2} Dirac RPM can be taken to use $\FrQ(\mathbb{R}^d, N) = \mathbb{R}^{N \, d}$ variables.\footnote{Or, just as well, 
%OOOOOOOOOOOOOOOOOOOOOOOOOOOOOOOOOOOOOOOOOOOOOOOOOOOOOOOOOOOOOOOOOOOOOOOOOOOOOOOOOOOOOOOOOOOOOOOOOOOOOOOOOOOOOOOOOOOOOOOOOOOOOOOOOOOOOOOOOOOOOOOOOOOOOOOOOOOOOOOOOOOOOOOOOOOOOOOOOOOO
relative space $\Frr(\mathbb{R}^d, N) = \mathbb{R}^{n \, d}$ variables, using Article II's observations.}
%OOOOOOOOOOOOOOOOOOOOOOOOOOOOOOOOOOOOOOOOOOOOOOOOOOOOOOOOOOOOOOOOOOOOOOOOOOOOOOOOOOOOOOOOOOOOOOOOOOOOOOOOOOOOOOOOOOOOOOOOOOOOOOOOOOOOOOOOOOOOOOOOOOOOOOOOOOOOOOOOOOOOOOOOOOOOOOOOOOOO

\m 

\n Reduced RPM uses shape(-and-scale) variables, also provided in Article II in some simple cases and more comprehensively in Part III of \cite{ABook}.

\m 

\n{\bf Example 3}  For minisuperspace, there is no reduced--Dirac distinction.  
That scalefactor 
\be 
a \geq 0
\ee 
is moreover significant \cite{BI75}, for reasons already modelled by the quantization of the real half-line $\mathbb{R}_+$ 
being substantially distinct \cite{I84} from that of the whole real line $\mathbb{R}$. 
While on the one hand this kind of effect\footnote{\cite{AKin} and Part III of \cite{ABook} 
%OOOOOOOOOOOOOOOOOOOOOOOOOOOOOOOOOOOOOOOOOOOOOOOOOOOOOOOOOOOOOOOOOOOOOOOOOOOOOOOOOOOOOOOOOOOOOOOOOOOOOOOOOOOOOOOOOOOOOOOOOOOOOOOOOOOOOOOOOOOOOOOOOOOOOOOOOOOOOOOOOOOOOOOOOOOOOOOOOOOO
consider RPM and SIC counterparts of this effect.} 
%OOOOOOOOOOOOOOOOOOOOOOOOOOOOOOOOOOOOOOOOOOOOOOOOOOOOOOOOOOOOOOOOOOOOOOOOOOOOOOOOOOOOOOOOOOOOOOOOOOOOOOOOOOOOOOOOOOOOOOOOOOOOOOOOOOOOOOOOOOOOOOOOOOOOOOOOOOOOOOOOOOOOOOOOOOOOOOOOOOOO
is global in contrast to the current Series' local treatment, almost all the literature errs at this point. 

\m 

\n{\bf Example 4}  The above consideration moreover affects GR-as-Geometrodynamics as well, via the classical inequality
\be 
\mbox{det} \, \bh \m > \m 0 \m . 
\ee 
This is incorporated by Affine Quantum Geometrodynamics' \cite{IK84, Klauder06} choice of kinematical quantization but not by the plain-or-unqualified Quantum Geometrodynamics 
which is far more commonly used in the literature. 

\m 

\n Full GR requires the Dirac approach out of not having a specific analytic general solution to classical reduction. 
This necessitates Dirac-quantized model arenas which as RPM treated this way, for all that in this case the reduced approach is also amenable 
(and thus further available for useful comparison). 

\m 

\n{\bf Structure 2} Kinematical quantization involves moreover passage from classical Poisson (or Dirac) brackets to quantum commutators. 
Our choice of operators then needing to close under the latter brackets. 
Affine Quantum Geometrodynamics is then an example of passage from a Poisson brackets algebra to an {\sl inequivalent} quantum commutator algebra.  
See Chapters 39 to 43 of \cite{ABook} for further details of passage between types of brackets and phenomena thereby exhibited.

%====================================================================================================================================================================================
%=========================================================================================================================================================================================
\section{Quantum Temporal Relationalism (aspect 0a)}\label{Q-TR}
%=========================================================================================================================================================================================
%====================================================================================================================================================================================

I proceed via giving a {\it TRiCQT} \cite{ABook}: a TRi variant of Isham's approach to Canonical Quantum Theory \cite{I84}.  
This clearly indicates a Canonical Quantum Theory sequel to TRiPoD.  
Kinematical Quantization is already-TRi, so TRiCQT first becomes significant in the current Section's material.

%=========================================================================================================================================================================================
\subsection{Quantum Chronos constraint}\label{QWE}
%=========================================================================================================================================================================================

GR gives rise to a stationary wave equation -- the {\it Wheeler--DeWitt equation} $\widehat{\scH} \, \Psi = 0$ -- 
in a context in which simpler theories' QM gives a time-dependent wave equation.   
These authors called this the {\it Einstein--Schr\"{o}dinger equation}; a more detailed form for this is 
$$
\widehat{\scH} \, \Psi  \:= - \hbar^2 ` \Box_{\sbM} - \xi \, {\cal R}_{\sbM}(\uix; \bh]\mbox{'}\,\Psi 
-\sqrt{\mh} \, {\cal R} \, \Psi + 2\sqrt{\mh} \, \slLambda \, \Psi + \widehat\scH^{\mbox{\scriptsize matter}}\Psi
$$
\be 
\:=  - \hbar^2  `  \frac{1}{\sqrt{\mM}}  \frac{\updelta}{\updelta \u{\u{\bh}}}
\left\{
\sqrt{\mM} \, \u{\u{\u{\u{\bN}}}}  \frac{  \updelta\Psi  }{  \updelta \u{\u{\bh}}  }
\right\} 
- \xi \,{\cal R}_{\sbM}(\uix; \bh]\mbox{'}\,\Psi 
-\sqrt{\mh} \, {\cal R} \, \Psi + 2\sqrt{\mh} \, \slLambda \, \Psi + \widehat\scH^{\mbox{\scriptsize matter}}\Psi  = 0   \m . 
\label{WDE2}     
\eeq
Solving this is termed {\it dynamical quantization} \cite{I84}.  
` \m ' here incorporates the following subtleties.  

\m 

\n 1) This formula has various well-definedness issues as further outlined in Part III of \cite{ABook}.
In any case, these are absent from finite models such as Minisuperspace (\ref{MSS-WDE}). 

\m 

\n 2) Operator-ordering issues which still partly remain for finite models.   

\m 

\n{\bf Example 1} Minisuperspace retains a counterpart of the Wheeler--DeWitt equation.  
Indeed, Misner's introduction of Minisuperspace \cite{mini, Magic} was largely motivated by the wish to study more tractable versions of (\ref{WDE2}). 
His original models \cite{mini, Magic} were isotropic and anisotropic, in each case without fundamental matter.
For this Series of Articles's main Minisuperspace model, with single minimally-coupled scalar field matter,   
the Wheeler--DeWitt equation reads 
\beq
\widehat{\scH}_{\sm\si\sn\si} \, \Psi  \:=  - \hbar^2 \Box_{\sbM} - \mbox{exp}(6 \, \slOmega)\{\mbox{exp}(\slOmega) - 2 \, \slLambda - V(\phi) \}\Psi 
                                       \:=    \hbar^2 \{\pa_{\slOmega}^2 - \pa^2_{\phi}\}\Psi - \mbox{exp}(6 \, \slOmega)\{\mbox{exp}(\slOmega) - 2 \, \slLambda - V(\phi) \}\Psi  \m .
\label{MSS-WDE}
\eeq
\n{\bf Example 2} The unreduced Euclidean RPM model arena counterpart is 
\beq
E \, \Psi  =  \widehat{\scE} \, \Psi  
           =  - \frac{\hbar^2}{2} \triangle_{\sbiM} \Psi + V \, \Psi                                \m . 
\eeq
\n{\bf Remark 1} Compared to Minisuperspace this model permits tractable study of accompanying linear quantum constraints and of structure formation.

%=========================================================================================================================================================================================
\subsection{The Frozen Formalism Problem (facet 1)}\label{FFP}
%=========================================================================================================================================================================================

The Wheeler--DeWitt equation is one of the places in which the most well-known Problem of Time facet -- the Frozen Formalism Problem -- appears, 
since one would be expecting a time-dependent wave equation at this point (for some notion of time).

\m 

\n The most well-known {\it (Schr\"{o}dinger-Picture) Quantum Frozen Formalism Problem} 
arises from elevating a quadratic constraint equation $\Quad$, encompassing both GR's $\scH$ and RPM's $\scE$, to a quantum equation
\beq
\widehat{\Quad} \, | \, \Psi \, \rangle = 0  \m . 
\label{E=0-TISE} 
\eeq
Here, $\Psi$ is the {\it quantum wavefunction of the} (model) {\it universe}. 

\m 

\n See (\ref{WDE2}) for the detailed form of the GR case of this equation: the so-called {\it Wheeler--DeWitt equation} \cite{Battelle, DeWitt67}.     
This is often viewed as the $E = 0$ case of a time-independent Schr\"{o}dinger equation 
\be 
\widehat{H} \Psi = E \Psi \m :
\ee
a stationary alias timeless or frozen quantum wave equation.  
This occurs in a place in which one would expect a time-dependent quantum wave equation such as the time-dependent Schr\"{o}dinger equation 
\beq
i \, \hbar \, \frac{\pa\uppsi}{\pa t} = \widehat{H} \, \uppsi                                                               \m ,   
\label{TDSE}
\eeq  
the Klein--Gordon equation
\beq
\hbar^2 \Box \phi = m^2  \phi \mbox{ } \mbox{ for wave operator } \mbox{ } \mbox{ }  \Box := \pa_t^2 + \triangle  \m ,  
\label{KGE}
\eeq
or the Dirac equation. 
\be 
\{ i \, \hbar \, \gamma^{\mu}\mbox{}_{B}\mbox{}^{A} \pa_{\mu} - m  \, \delta_{B}\mbox{}^{A}\}\uppsi^{B} = 0               \m , 
\label{Dirac-Eq}
\ee
for $\u{\u{\u{\gamma}}}$ the vector of Dirac matrices.  
%
% Are all in $c = 1$ units.
% 
\n On occasion, this has been interpreted at face value as a Fully Timeless Worldview arising from attempting to combine GR and Quantum Theory.   
See however Parts II and III of \cite{ABook} for further interpretations and means of bypassing such an equation arising in the first place.  

\m

\n (\ref{TDSE}) is presented above in the finite-theory case for simplicity (so its given form includes just the Minisuperspace subcase of GR).
The field-theoretic counterpart of (\ref{E=0-TISE}) contains, 
\be 
\mbox{in place of a partial derivative } \m \frac{\pa}{\pa \bQ}                   \mma 
\mbox{a functional derivative } \m          \frac{\updelta}{\updelta \bh(\u{x})}  \m . 
\ee
The Wheeler--DeWitt equation arises regardless of whether from ADM's scheme that presupposes and  subsequently splits spacetime, 
or as an equation of time $\Chronos$ from implementing Temporal Relationalism as per above.
Moreover, from the latter perspective, the Frozen Formalism Problem already features at the classical level for the Universe as a whole;  
its being manifested at the quantum level is then less surprising. 

\m  

\n One of this Series of Articles's main points is that the Wheeler--DeWitt equation of GR, $\widehat{\scH} \, \Psi = 0$, 
can be traced back not only to the classical Hamiltonian constraint $\scH$ 
but furthermore to               Temporal Relationalism.

\m 

\n Temporal Relationalism provides constraints for the range of formulations of theories which implement this principle.
These constraints are interpreted as equations of time, denoted in general by $\Chronos$;
this interpretation provides a classical emergent Machian time resolution of the ab initio timelessness of these formulations. 
Both $\scH$ and $\scE$ can be taken to arise in this manner.
Moreover, in each theory which possesses a $\Chronos$, this leads to an also apparently frozen quantum wave equation 
\be 
\widehat{\Chronos} \, \Psi = 0  \m .
\ee

%==================================================================================================================================================================================
\subsection{Inner Product Problem alias Hilbert Space Problem}\label{IPP} 
%==================================================================================================================================================================================

In Quantum Theory, the wave equation does not suffice to obtain physical answers, since these are of the form 
\be 
\langle \, \uppsi_1 \, | \, \widehat{O} \, | \, \uppsi_2 \, \rangle \m ,
\ee 
so an inner product input is also required. 

\m 

\n The Schr\"{o}dinger inner product serves this purpose in Ordinary QM.  
Klein--Gordon Theory has its own distinct inner product.    
Recollect that a Schr\"{o}dinger inner product will not do in this setting because $\mathbb{M}^4$ is indefinite, which argument carries over to GR's 
$\Riem(\bupSigma)$ and minisuperspace $\Mini(\bupSigma)$ as well.  

\m 

\n In the Quantum Gravity of the above Wheeler--DeWitt equation, the inner product is, at least prima facie, of Klein--Gordon type:  
\be 
\langle \,\uppsi_1[\bh] \, | \, \uppsi_2[\bh] \, \rangle   \es  \frac{1}{2 \, i} \prod\mbox{}_{\mbox{}_{\mbox{\scriptsize $x \in \sbupSigma$}}} \int 
\left( 
\d\bupSigma \mma  
\uppsi_1[\bh]\stackrel{\longleftrightarrow}{\frac{\updelta}{\delta \bh }} \uppsi_2[\bh]
\right)_{\sbN}                                                                                                                                      \es 
\  \frac{1}{2 \, i} \prod\mbox{}_{\mbox{}_{\mbox{\scriptsize $x \in \sbupSigma$}}} 
                                                             \int \d\bupSigma_{ij} \mN^{ijkl}(\bh)
\left\{
\uppsi_1[\bh]\stackrel{\longleftrightarrow}{\frac{\updelta}{\delta \mh_{kl} }} \uppsi_2[\bh]
\right\}                                                                                                                                              \m .
\label{KG-IP2}
\ee
This however runs into further technical problems as outlined in Chapter 11 of \cite{ABook}.    

\m 

\n{\bf Example 1)} Minisuperspace retains a Klein--Gordon inner product.   

\m 

\n{\bf Example 2)} Euclidean RPM, on the other hand, has a natural associated Schr\"{o}dinger Inner Product, like ordinary QM does. 
This is on account of each of these having a positive-definite metric on $\FrQ$, 
yielding a positive-definite inner product for which a Schr\"{o}dinger interpretation is appropriate.  

\m  

\n{\bf Remark 1} The Inner Product Problem is a temporal issue -- a subfacet of the Frozen Formalism Problem -- 
due to the ties between inner products, conservation of probability and unitary evolution outlined in Chapter 5 of \cite{ABook}.

\m 

\n{\bf Aside 1} Let us point to Quantum Theory including further objects such as quantum operators and path integrals; 
see Part III of \cite{ABook} for how these can be cast into TRi form.

%=========================================================================================================================================================================================
\subsection{Semiclassical Machian emergent time}\label{t-sem}
%=========================================================================================================================================================================================

\n The classical $t^{\se\sm(\sJ)}$ does not itself resolve the quantum-level Frozen Formalism Problem, 
nor does it in any other way directly give quantum equations that are distinct from the usual ones.   
One can view it, rather, as an object that already feature at the classical level that is subsequently to be recovered by a more bottom-up approach 
at the quantum level. 

\m  

\n We instead use semiclassical Machian emergent time $t^{\sss\se\sm}$, 
in situations in which there are slow, heavy `$h$'  variables that provide an approximate timestandard 
with respect to  which the other fast, light `$l$' degrees of freedom evolve \cite{HallHaw, K92, KieferBook}.  
This occurs e.g.\ in the Semiclassical Approach for SIC \cite{HallHaw, SIC-1, SIC-2}, $h$ is scale (and homogeneous matter modes) 
%                                                                         and $l$ are one or both of small anisotropies or small inhomogeneities. 
%
The Semiclassical Approach consists of the following steps. 

\m

\n{\bf Step i)} Make the Born--Oppenheimer ansatz 
\beq
\Psi(h, l)  =  \uppsi(h) \, | \, \chi(h, l) \, \rangle \m ,  
\label{BO}
\eeq 
followed by the WKB ansatz
\beq
\uppsi(h) = \mbox{exp}\left(\frac{i\,S(h)}{\hbar}\right) \m .
\label{WKB}
\eeq 
Each of these is accompanied by a suite of approximations, detailed in Section \ref{t-sem}.  

\m 

\n{\bf Step ii)} We form the $h$-equation
\beq
\langle \,\chi| \widehat{\Quad} \, \Psi  =  0  \m .
\eeq 
To first approximation, this yields a Hamilton--Jacobi equation \cite{Goldstein, Lanczos},\f{For simplicity,
%OOOOOOOOOOOOOOOOOOOOOOOOOOOOOOOOOOOOOOOOOOOOOOOOOOOOOOOOOOOOOOOOOOOOOOOOOOOOOOOOOOOOOOOOOOOOOOOOOOOOOOOOOOOOOOOOOOOOOOOOOOOOOOOOOOOOOOOOOOOOOOOOOOOOOOOOOOOOOOOOOOOOOOOOOOOOOOOOOOOOOOOOOO
we present the rest of this Section for Mechanics with one $h$ degree of freedom; see \cite{FileR, QuadII} for consideration of multiple such and other generalizations. }
%OOOOOOOOOOOOOOOOOOOOOOOOOOOOOOOOOOOOOOOOOOOOOOOOOOOOOOOOOOOOOOOOOOOOOOOOOOOOOOOOOOOOOOOOOOOOOOOOOOOOOOOOOOOOOOOOOOOOOOOOOOOOOOOOOOOOOOOOOOOOOOOOOOOOOOOOOOOOOOOOOOOOOOOOOOOOOOOOOOOOOOOOOO
\beq
\left\{\frac{\pa S}{\pa h}\right\}^2  \es  2\{E - V(h)\}  \m ,
\eeq
where $V(h)$ is the $h$-part of the potential. 
Furthermore, one way of solving this is for an approximate semiclassical emergent time $t^{\sss\se\sm}(h)$. 

\m 

\n{\bf Step Small} We next consider the $l$-equation 
\beq
\{1 - | \, \chi \, \rangle\langle \, \chi \, |\} \widehat{\Quad} \,\Psi = 0 \m . 
\label{FLUC}
\eeq 
In this initial form, this is a fluctuation equation.
Moreover, it can be recast -- modulo some more approximations -- into an emergent-WKB-time-dependent Schr\"{o}dinger equation for the $l$-degrees of freedom. 
For instance, for Mechanics
\beq
i \, \hbar \frac{\pa \, | \, \chi \, \rangle}{\pa t^{\sss\se\sm}}  \es  \widehat\scE_{l} \, | \, \chi \, \rangle \m .
\label{TDSE2}
\eeq
The emergent-time-dependent left hand side arises from the cross-term $\pa_{h} \, | \, \chi \, \rangle \, \pa_{h}\uppsi$.  
$\widehat\scE_{l}$ is here the remaining piece of $\widehat\scE$, which plays the role of Hamiltonian for the $l$-subsystem.  

\m 

\n{\bf Step iv)} In this Series of Articles's main approach, $\Quad$ arises as an equation of time $\Chronos$.
$t^{\sss\se\sm}$ can furthermore be interpreted as \cite{FileR, ABook} a semiclassical Machian emergent time (whether for the above model arena or for GR Quantum Cosmology). 

\m 

\n To zeroth order in $l$, this and the classical 
\be 
t^{\se\sm(\sJ)-0}  =  F(h, \d h)  =  t^{\sss\se\sm(\sJ)-0}  \m .
\ee   
This however fails to be Machian in the sense of not permitting either classical or semiclassical $l$-change to contribute.   

\m 

\n And yet, to first order in $l$ 
\beq
t^{\sss\se\sm-1}   =  F[h, l, \d h, |\chi(l, h) \, \rangle]  \m ,
\eeq 
which is now clearly distinct from the classical $h$--$l$ expansion's  
\beq
t^{\se\sm(\sJ)-1}  =  F[h, l, \d h, \d l]                    \m .
\eeq
This pairing of the previously known semiclassical emergent time 
and the classical Machian emergent time is new to the Relational Approach, as is the Machian reinterpretation of the former.  
That emergent Machian time needs to be `found afresh' at the quantum level, rather than continuing to use the classical $t^{\se\sm}$, 
is moreover  in accord with the 
`all changes have an opportunity to contribute' implementation of Mach's Time Principle.   
This is for the furtherly Machian reason that abstracting GLET from STLRC 
is sensitive to partial or total replacement of classical change by {\sl quantum change} being given the opportunity to contribute.

%=========================================================================================================================================================================================
\subsection{Commentary}
%=========================================================================================================================================================================================

That many other strategies have been attempted at this point -- internal time, hidden time, matter time, unimodular time, timelessness, histories theory... -- 
is reviewed in \cite{K92, I93, ABook}.
The one we provide is the only one known to work to date in combination with addressing all other facets of `a local Problem of Time'.

\m 

\n Justifying a semiclassical regime is a further matter, along the following lines. 

\m 

\n{\bf Remark 1)} It is a matter decoupled from finding a local resolution of Problem of Time within classical and semiclassical setting.

\m 

\n As per Part III of \cite{ABook},  
the above derivation of a time-dependent Schr\"{o}dinger equation ceases to function if the WKB scheme (ansatz and approximation).  
Moreover, in the quantum-cosmological context, the WKB scheme is not known to be a particularly strongly supported ansatz and approximation to make 
(see Section \ref{t-sem}).      
This Series props this up by combination with further Problem of Time strategies, which need to be individually developed from the classical level upwards. 
 
\m 

\n{\bf Remark 2)} It is a matter that the Combined Approach -- elements of semiclassical, histories and timeless approaches -- does a better job of \cite{ABook}.
While most of the supports between these are quantum-specific, there is no problem in setting up a classical precursor of this, 
ultimately a TRi version of the previous combined approach of Halliwell \cite{H99, H03, H09}.  

\m 

\n{\bf Remark 3)} The idea that semiclassical Machian time is not classical Machian time due to semiclassical change needing to be given an opportunity to contribute 
gives a fair indication of full QM with full QM change also serving as basis of abstraction of a fully quantum time. 
Machian emergent time is thus expected to be a universally applicable concept rather than just a classical and semiclassical one. 

\m 

\n{\bf Remark 4)} Observations of Quantum Gravity in the forseeable future (40-year timescale, at least) are likely to be at most semiclassical. 
The SIC model is specifically for cosmological such \cite{HH83, HallHaw}; the sensible next move would then be to set up semiclassical Machian emergent time 
in the black hole arena, rather than for 'full Quantum Gravity'.  
So there is an element of realistic sufficiency to working semiclassically, for all that the principles in use at least conceptually transcend to this regime as per 3).

%=========================================================================================================================================================================================
%=========================================================================================================================================================================================
\section{Quantum Configurational Relationalism (aspect 0b)}\label{Q-CR}
%=========================================================================================================================================================================================
%=========================================================================================================================================================================================

We now consider Quantization for a physical theory subject to a group $\lFrg$ of physically meaningless transformations.
Ab initio, there are two ways of quantizing such a theory.  

\m 

\n{\bf Strategy i) Reduced Quantization}. 
Here one first reduces out the constraints corresponding to $\lFrg$ at the classical level, and then one quantizes.  

\m 

\n{\bf Strategy ii) Dirac Quantization}. 
Here one quantizes first.
The constraints corresponding to $\lFrg$ are now promoted to further quantum wave equations    
\beq
\widehat{\bFlin} \, \Psi = 0  \m , 
\eeq
which are then reduced out -- solved -- at the quantum level.

\m

\n{\bf Remark 1} That Quantum Theory is capable of discarding a physically-accepted $\lFrg$ may affect Reduced Quantization. 
Thereby, classical ability to reduce out the classical $\lFrg$ (`Best Matching')
does not necessarily imply an $\lFrg$-free quantum system (see Part III of \cite{ABook} for details).  

\m 

\n{\bf Remark 2} The indirect $\lFrg$-act, $\lFrg$-all method moreover continues to be applicable at the quantum level. 
This could, firstly, be as a means of formulating Dirac Quantization.
Secondly, it could be as an indirect means of expressing all subsequent objects required by one's theory if neither Stratgies 1) or 2) can be culminated.

%===================================================================================================================================================================
\subsection{Dirac quantization approach}
%===================================================================================================================================================================

Let us promote the classical first-class constraints arising from Configurational Relationalism and Temporal Relationaism to quantum constraints, 
\be 
\Chronos  \m \longrightarrow \m  \widehat{\Chronos}    \m .
\ee
\be 
\bFlin     \m \longrightarrow \m  \widehat{\bFlin}     \m ,
\ee
These act on the {\it wavefunction of the Universe} $\Psi$. 

\m 

\n{\bf Example 1} full GR, the quantum Hamiltonian constraint is accompanied by the quantum GR momentum constraint 
\beq
0  \es  \widehat{\u{\sbcM}} \Psi  
\ee
or, in components, 
\beq
0  \es  2 \, i \, \hbar \, \mh_{ik} {\cal D}_j \frac{\updelta}{\updelta \mh_{jk}}\Psi + \widehat\scM_i^{\mbox{\scriptsize matter}}\Psi   \m , 
\label{Q-Mom}
\eeq
\n{\bf Example 2} $\u{\sbcM}$ is of course absent for our Minisuperspace model. 

\m 

\n{\bf Example 3} For Euclidean RPM,   
\be
\hat{\u{\sbcP}} \,  \Psi \es \mbox{$\frac{\hbar}{i}$} \sumIN  \frac{\pa\Psi}{\pa \u{q}^{I}} \es 0       \m , 
\label{P-hat}
\ee 
\beq
\widehat{\u{\sbcL}} \, \Psi \es \mbox{$\frac{\hbar}{i}$} \sumIN \u{q}^I \cr \frac{\pa \Psi}{\pa \u{p}^I}  = 0           \m ,
\label{L-hat}
\eeq
\be 
\hat{\w{E}}\Psi  \es  -\mbox{$\frac{\hbar^2}{2}$} \triangle_{\mathbb{R}^{nd}} \Psi + V(\mbox{\bf--}\bullet\mbox{\bf--})\Psi 
                         \es    E \, \Psi                                                                      \m . 
\ee

\n Ditching the zero momentum constraint $\u{\scP}$, and using Sec II.6.2's $\urho^A$, 
\beq
\widehat{\scbL} \, \Psi  \es \mbox{$\frac{\hbar}{i}$} \sumAn \urho^A \cr \frac{\pa \Psi}{\pa\urho^A}  = 0
\label{L-hat-2}
\eeq
holds.    
This implies that $\Psi = \Psi(\mbox{\bf--}\bullet\mbox{\bf--})$.  
Then finally   
\beq
\widehat{\scE} \, \Psi = -\mbox{$\frac{\hbar^2}{2}$} \triangle_{\mathbb{R}^{nd}} \Psi + V(\mbox{\bf--}\bullet\mbox{\bf--})\Psi  
\es  E \, \Psi \mbox{ } . 
\label{E-hat}
\eeq

%===================================================================================================================================================================
\subsection{Reduced quantization approach}
%===================================================================================================================================================================

The classically-reducible scaled-RPM series' \index{time-independent Schr\"{o}dinger equation}time-independent Schr\"{o}dinger equations are therefore encapsulated by the $d = 1, 2$ 
cases of 
\beq
-\hbar^2 \left\{ \pa^2_{\rho} + k(N, d)\rho^{-1}\pa_{\rho} + \rho^{-2}\triangle_{\sfS(N, d)}\right\} \Psi  \es  
2\{E_{\sU\sn\si} - V(\rho, \bis)\}\Psi \mbox{ } . 
\label{Gilthoniel}
\eeq
(\ref{Gilthoniel}) furthermore separates into scale and shape parts for a number of suitable $V$
See Part III of \cite{ABook} for a more specifically-motivated operator ordering of this.  
$\bis$ here denote shape variables.

%==================================================================================================================================================================================
%==================================================================================================================================================================================
\section{Quantum Constraint Closure (aspect 1) including the Functional Evolution Problem (facet 1)}\label{Q-CC}
%==================================================================================================================================================================================
%==================================================================================================================================================================================

Commutator brackets play an even more central role in Quantum Theory than Poisson brackets did at the classical level.
Moreover, the quantum notion of {\sl equal-time} commutation relations poses significant difficulties in the context of GR. 
This is due to Ordinary Quantum Theory's `equal-time' notion carrying connotations of there being a unique preassigned time, which does not fit GR's conception of time.

\m 

\n{\bf Structure 1} A first instance of equal-time commutation relations occurs in Kinematical Quantization (Section \ref{Kin-Quant}).  

\m 

\n{\bf Remark 1} At the quantum level, constraints take the form of operator-valued equations. 
Moreover, passage from classical to quantum constraints is subject to operator-ordering ambiguities and well-definedness issues. 
One is then also to consider commutator brackets between these quantum constraints. 
For sure, algebraic closure of constraints is not automatically guaranteed in postulating the form these are to take at the quantum level: 
\beq
\widehat{\bscC} \, \Psi                                                                                  =  0  \m \centernot{\Rightarrow} \m \m  
\mbox{\bf [} \, \widehat{\bscC}    \mbox{\bf ,} \, \widehat{\bscC}  \, \mbox{\bf ]} \, \Psi  =  0  \m . 
\eeq 
Or perhaps the Lie-weak 
\beq
\widehat{\bscC} \, \Psi                                                                                  =  0  \m \centernot{\Rightarrow} \m \m  
\mbox{\bf [} \, \widehat{\bscC}    \mbox{\bf ,} \, \widehat{\bscC}  \, \mbox{\bf ]} \, \Psi  \es   \u{\u{\u{\biC}}} \cdot \widehat{\bscC} \, \Psi  \m .    
\eeq 
\n{\bf Remark 2} Commutator bracket algebraic structures are furthermore not in general isomorphic to their classical Poisson brackets antecedents or approximands.  
Section \ref{Kin-Quant} outlines the topological underpinnings of this discrepancy \cite{I84}.
For constraint algebraic structures, this means that the quantum version is not necessarily isomorphic to the classical one. 

\m 

\n{\bf Remark 3} One consequence of this is that algebraic closure of classical constraints does not imply an isomorphic algebraic closure of quantum constraints, 
nor indeed of any other kind of quantum-level closure:  
\beq
\mbox{\bf \{} \,   \bscC    \mbox{\bf ,} \, \bscC  \,  \mbox{\bf \}}                   \approx  0 
\m \centernot{\Rightarrow}  \m \m 
\mbox{\bf [}  \,  \widehat\bscC    \mbox{\bf ,} \, \widehat\bscC  \,  \mbox{\bf ]} \, \Psi     \es   0  \m . 
\label{C-C-2}
\eeq
Furthermore, the latter set of constraints in general requires a distinct indexing set in place of $\fC$.  
Anomalies are one manifestation of non-closure that is well-known throughout Theoretical Physics \cite{Dirac, AW84, Bertlmann}.
It additionally a means by which a classically accepted $\lFrg$ may need to be replaced by a distinct $\lFrg_{\sQ\sM}$ at the quantum level.
While not all anomalies involve time or frame, a subset of them do, and these then form part of the Problem of Time. 
These issues are further developed in Section \ref{Q-CC}.

\m 

\n{\bf Remark 4} Breakdown of the closure of the constraint algebraic structure at the quantum level was termed the {\it Functional Evolution Problem} in \cite{K92, I93}. 
However, `functional' carries field-theoretic connotations -- it is the type of derivative that features in the field-theoretic form of the problem.  
Articles VI and VII and iron this out in species-neutral terms.  
Furthermore, to additionally include the classical case and maximally clarify the nature of this problem, 
we consider it better to refer to this facet as the {\it Constraint Closure Problem}.  

\m 

\n{\bf Remark 5} Many approaches to Quantization (see \cite{I84} or Part III of \cite{ABook}) are of at most limited value in the case of GR. 
This is due to the classical GR constraints forming the Dirac algebroid, whereas many an established approach to Quantization can only cope with Lie algebras.  

\m 

\n{\bf Structure 2} We next need to check the quantum first-class constraints $\widehat{\bscF}$ close,  
\be 
\mbox{\bf [} \, \widehat{\bscF} \mbox{\bf ,} \, \widehat{\bscF} \, \mbox{\bf ]} \, \Psi  =  0  \m .
\ee 
\n{\bf Structure 3} The Configurational to Temporal Relational split version of this is 
\be 
\mbox{\bf [}  \, \widehat{\bFlin} \mbox{\bf ,} \, \widehat{\bFlin} \, \mbox{\bf ]} \, \Psi = 0                       \m ,  
\ee
\be 
\mbox{\bf [} \, \widehat{\bFlin} \mbox{\bf ,} \,  \widehat{\Chronos} \, \mbox{\bf ]} \, \Psi = 0                    \m , 
\ee
\be 
\mbox{\bf [} \, \widehat{\Chronos} \mbox{\bf ,} \, \widehat{\Chronos} \, \mbox{\bf ]} \, \, \Psi = 0                \m .  
\ee

%==================================================================================================================================================================================
%==================================================================================================================================================================================
\section{Quantum Assignment of Observables (aspect 2 reopened)}\label{Q-EitoO}
%==================================================================================================================================================================================
%==================================================================================================================================================================================

\n{\bf Structure 1} For systems which additionally possess quantum constraints, 
the $\widehat{\beables}$ are additionally to form zero quantum commutators with the quantum constraint operators.  

\m 

\n{\bf Example 1} 
\beq
\mbox{Quantum Dirac observables: $\widehat{\Dirac}$ such that } \m \mbox{\bf [} \, \widehat{\sbcF} \, \mbox{\bf ,}  \m \widehat{\Dirac} \, \mbox{\bf ]} \, \Psi = 0 \m ,
\label{QDB}
\eeq
\n{\bf Example 2}
\beq
\mbox{quantum \K observables: $\widehat{\Kuchar}$ such that } \m    \mbox{\bf [} \, \widehat{\bFlin}\mbox{\bf ,} \, \m \widehat{\Kuchar} \, \mbox{\bf ]} \, \Psi = 0      \m .
\label{QKB}
\ee 
\n{\bf Example 3}
\beq
\mbox{quantum $\lFrg$-observables: $\widehat{\gauge}$ such that } 
                                                             \m \mbox{\bf [} \, \widehat{\bGauge} \mbox{\bf ,} \, \m \widehat{\gauge} \, \mbox{\bf ]} \, \Psi = 0          \m .
\label{QGB}
\ee
These need not coincide with the previous. 

\m 

\n{\bf Example 4} A further notion of 
\beq
\mbox{quantum Chronos observables: $\widehat{\chronos}$ such that } \m \mbox{\bf [} \, \widehat{\Chronos}  \mbox{\bf ,} \,   \widehat{\chronos} \,  \mbox{\bf ]} \, \Psi = 0 
\label{QCB}
\eeq
exists in cases for which $\widehat{\Chronos}$ closes as a subalgebraic structure. 

\m 

\n{\bf Remark 1} We could also entertain the possibility of taking each of these to be a weak brackets relation, 
\be 
\mbox{\bf [} \, \widehat{\u{\sbcC}} \mbox{\bf ,} \, \widehat{\u{\sbiB}} \, \mbox{\bf ]} \, \Psi  \es  \u{\u{\u{\biE}}} \cdot \widehat{\sbcC} \, \Psi \m .   
\ee 
\n For, the associated brackets might be expected to carry the same notion of equality as the initial constraints brackets algebraic structure itslef...

\m 

\n{\bf Structure 2} We need to upgrade from $\hat{\sbiU}$ to $\hat{\sbiD}$ observables: 
quantities commuting with all the surviving first-class quantum constraints $\widehat{\sbcC}$
In this way, quantum observables are `found afresh'. 
This is as opposed to trying to promote classical observables, which would face three obstacles. 

\m 

\n{\bf Obstacle 1)} The change from Poisson brackets to commutators. 

\m 

\n{\bf Obstacle 2)} The restrictions already imposed by kinematical quantization. 

\m 

\n{\bf Obstacle 3)} The promotion of classical constraints $\sbcC$ to their quantized forms $\widehat{\sbcC}$.  

\m 

\n The {\bf Problem of Quantum Observables} is that it is hard to come up with a sufficient set of these for Quantum Gravitational theories. 

\m 

\n{\bf Remark 1} In the Heisenberg picture of QM, an apparent manifestation of frozenness is, rather,  
\beq
\mbox{\bf [} \, \widehat{H} \mbox{\bf ,} \, \widehat{\beables} \, \mbox{\bf ]} \Psi = 0
\eeq
with similar connotations to its classical Poisson brackets antecedent (III.80).

%===================================================================================================================================================================================
%=========================================================================================================================================================================================
\section{Quantum Constructability of Spacetime from Space (aspect A)}\label{Q-SCP}
%=========================================================================================================================================================================================
%===================================================================================================================================================================================

{\bf Wheeler's first argument} \cite{Battelle} Fluctuations of the dynamical entities are inevitable at the quantum level. 

\m 

\n For GR, moreover, these are fluctuations of 3-geometry.  

\m 

\n These fluctuating geometries are additionally far too numerous to be embeddable within a single spacetime.  
Consequently, the beautiful geometrical manner in which that classical GR manages to be Refoliation Invariant breaks down at the quantum level.

\m 

\n{\bf Wheeler's second argument} \cite{Battelle, W79} gave the following additional argument.  
Precisely-known position $\u{q}$ and momentum $\u{p}$ for a particle are a classical concept tied to the notion of its worldline. 

\m 

\n This perspective breaks down in Quantum Theory, however, due to Heisenberg's Uncertainly Principle.  
Worldlines are here replaced by the more diffuse notion of wavepackets. 

\m 

\n In the case of GR, moreover, the Uncertainty Principle now applies to the quantum operator counterparts of $\bh$ and $\bp$. 
But, by formula (II.17), this means that $\bh$ and $\bK$ are not precisely known.   
The idea of embeddability of a 3-space with metric $\bh$ within a spacetime is consequently quantum-mechanically compromised.
Schematically,
\beq
\left( 
\stackrel{\mbox{ metric-level geometry}}
         {\mbox{ embedding data } \bh, \bK \m \mbox{ or } \bh, \bp  } 
\right) 
   \m \longrightarrow \m 
\left(  
\stackrel{\mbox{ operators } \widehat{\bh}, \, \widehat{\bp} \m \mbox{ subject to } }
         {\mbox{ Heisenberg's Uncertainty Principle}}  
\right) \m .  
\eeq
By these arguments, Geometrodynamics (or similar formulations) would be expected to take over from spacetime formulations at the quantum level.  

\m 

\n{\bf Remark 3} Recovering continuity in suitable limits, in approaches that treat space or spacetime as primarily discrete, is a further issue. 
This is not a given, since some approaches produce non-classical entities or too low a continuum dimension.  

\m 

\n One can next consider {\bf deformations at the semiclassical level} and see if {\bf semiclassical rigidity} \cite{ABook} ensues. 
Start e.g.\ with the geometrodynamics-assumed ($Diff(\bupSigma)$-invariant) but arbitrary supermetric ansatz 
\beq
\widehat{\scH}_{a, b, x, y} = -\frac{\hbar^2}{\sqrt{\mM_{x, y}}} \, \frac{\delta}{\delta \u{\u{\bh}}} 
\left\{ 
\sqrt{\mM_{x,y}} \u{\u{\u{\u{\bN}}}}\mbox{}_{x, y} \frac{\delta}{\delta \u{\u{\bh}}} 
\right\} \Psi 
- \hbar^2 \xi \, {\cal R}(\underline{x}; \mM_{x, y}] \Psi + \sqrt{\mh}\{ a + b {\cal R}(\underline{x}; \bh] \}\Psi
\mbox{ } , 
\eeq 
alongside the `momenta to the right' ordering of $\w{\u{\sbcM}}$, i.e.\ eq.\ (\ref{Q-Mom}).   
One is to compute the three commutator brackets formed by these, including upder Semiclassical Approach assumptions.

\m 

\n See also \cite{Bojo12} for further investigation of the semiclassical and quantum commutator bracket counterparts of the classical Dirac algebroid.

%=========================================================================================================================================================================================
%=========================================================================================================================================================================================
\section{Quantum Spacetime Relationalism (aspect 0$^{\prime}$)}\label{Q-STR}
%=========================================================================================================================================================================================
%=========================================================================================================================================================================================

Is spacetime -- or any of its aspects -- meaningful in Quantum Gravity? 

\m 

\n How does spacetime -- or any originally missing aspects thereof -- emerge in a suitable classical limit?  

\m 

\n Is there a notion in Quantum Gravity which resembles the causality of SR, QFT and GR? 
If so, which aspects of classical causality are retained as fundamental, and how do the others emerge in the classical limit?

\m 

\n Such questions lead to a Spacetime Relationalism versus Temporal-and-Configurational Relationalism debate. 
This in turn feeds into the following issues.   

\m 

\n{\bf Issue 1)} The quantum-level Feynman Path-Integral Approach versus Canonical Approach debate.  

\m 

\n{\bf Issue 2)} Consideration of whether quantum-level versions of Refoliation Invariance 
                                                   and Spacetime Constructability aspects of Background Independence are required. 
Some further specific quantum level issues about spacetime are as follows. 

\m  

\n{\bf Issue 3)} Whether a hypersurface is spacelike depends on the spacetime metric $\bg$.
In Quantum Gravity, however, this would be subject to quantum fluctuations \cite{I93}.  
In this way, the notion of `spacelike' would depend on the quantum state.

\m 

\n{\bf Issue 4)}  Relativity places importance upon labelling spacetime events by times and spatial frames of reference which are implemented by the deployment of physical clocks. 
What happens if one tries to model this using proper time at the quantum level \cite{I93}? 
Unfortunately, proper time intervals are built out of $\bg$, and thus are only meaningful after solving the equations of motion.
This is rendered yet more problematic by $\bg$'s quantum fluctuations.  

\m 

\n{\bf Issue 5)}  From a technical perspective, replacing SR spacetime's Poincar\'{e} group $Poin(4)$ by $Diff(\Frm)$ 
vastly complicates the Representation Theory involved.
Furthermore, the Representation Theory of the Dirac algebroid is {\sl even more} difficult than that of $Diff(\Frm)$.   

\m 

\n{\bf Issue 6)}  Finally, if one's approach attempts to combine spacetime and canonical concepts, there is additional interplay as e.g.\ outlined in Section \ref{Q-Fol}.

%=========================================================================================================================================================================================
\subsection{Path Integral Approaches}\label{PI-Intro}
%=========================================================================================================================================================================================

The Problem of Time facets do not take an entirely fixed form.  

\m 

\n On the one hand, if one splits spacetime and works with a 
Canonical Approach,  
the Frozen Formalism Problem,  
Inner Product Problem and 
Foliation Dependence Problem occur.

\m 

\n On the other hand, none of these occur if an unsplit spacetime formulation is used; Path Integral Approaches are along these lines. 

\m 

\n Because Path Integral Approaches are very successful in QFT, one might be tempted to approach Quantum Gravity in this way. 
These however face their own set of very major problems if one attempts to apply them to Gravity. 
In this way, the Canonical versus Path Integral dilemma amounts to choosing between two very different sets of sizeable problems.

\m 

\n{\bf Problem 1)} In place of an Inner Product Problem, we get a {\bf Measure Problem}.  
This is not limited to cases with timelike Killing vectors, but now requires $Diff(\Frm)$-invariance.  
Having an explicit $Diff(\Frm)$-invariant measure is a different -- but also considerable -- problem, and also a reason why the Measure Problem remains time-related.
More specifically, the Measure Problem is a further quantum-level part to Spacetime Relationalism. 
%
% Finally, this is the most direct reason for a self-contained book on the Problem of Time to carry an outline of what measures are (Appendix \ref{Fun-Meas-Prob}).   

\m 

\n{\bf Remark 1} While this is not necessarily a Background Dependent pursuit, this is not a Canonical Approach.  

\m 

\n Some further features of the Path Integral Approach for Gravitational Theories are as follows; 
consult Sec \ref{PI-Intro} for further details and references.

\m 

\n{\bf Problem 2)} It is, rather, a spacetime primary formulation.\index{spacetime!- primality}  
It is furthermore paths or histories that are now to be considered as primary.  

\m 

\n{\bf Problem 3)} The `wrong sign' of the GR action causes further problems for Path Integral Approaches.

\m 

\n{\bf Problem 4)} These problems are moreover ameliorated \cite{I81} by working in a Euclidean-signature sector, 
\beq
\langle \, \bh_{\si\sn}, t_{\si\sn} \, | \, \bh_{\sf\si\sn}, t_{\sf\si\sn}  \, \rangle  \es  
\int_{t_{\ti\tn}}^{t_{\tf\ti\tn}} \int_{\sbupSigma} \mD\mu[\bg] \, \mbox{exp}(- {\cal S}^{\sE\su\scc\sll\mbox{-}\sE\sH}_{\sG\sR}[\bg])   \m .
\eeq
Here, the integration is over all Euclidean-signature metric geometries on $\FrT \times \bupSigma$ in between, for $\FrT$ the time interval 
$[t_{\si\sn}, \, t_{\sf\si\sn}]$, and $\mD \mu$ denotes the measure.  

\m 

\n{\bf Problem 5)} However, GR's action is not positive-definite, which causes further unboundedness problems.   

\m 

\n{\bf Problem 6)} Moreover, Discrete Approaches to Quantum Gravity have better-defined path integrals.  
Such approaches started historically with Regge Calculus in the 1960s \cite{MTW}.  

\m 

\n{\bf Problem 7)} Time ordering and positive frequency impinge upon \cite{I85} the passage between Euclidean and Lorentzian sectors.  
At the perturbative level about flat spacetime this reduces to Wick's well-known QFT result.    
However, the curved spacetime counterpart of this picks up ambiguities as regards the choice of complex contour \cite{HL90}. 

\m 

\n{\bf Problem 8)} Let us finally point to some approaches which involve both Path Integrals and Canonical formalism. 
QFT can rely on the Canonical Approach for computing what its Feynman rules are. 
In this way, Ordinary Quantum Theory `in terms of path integrals' is in fact in some cases a combined Path Integral {\sl and} Canonical Approach.
Then in Quantum Gravity, such a combined scheme would pick up {\sl both} approaches' problems.

\m 

\n{\bf Remark 2} Upon passage to curved spacetime, what was flat spacetime's straightforward Wick rotation from imaginary time back to real time 
                                               becomes an ambiguity which is a further subfacet of the Problem of Time \cite{CarlipBook}.  
On these grounds, one might leave Path Integral Approaches to the QFT regime rather than entertaining their extension to Gravitational Theory, or not.

%===================================================================================================================================================================================
%===================================================================================================================================================================================
\section{Quantum Spacetime Closure (aspect 1$^{\prime}$)}
%===================================================================================================================================================================================
%===================================================================================================================================================================================

Suppose GR-like spacetime's emergence is a merely classical-level phenomenon. 
Then for some purposes it suffices to model its Relationalism classically, as per Article III. 

\m 
 
\n If there is already a semiclassical version of this emergence, 
then it is plausible that the physically irrelevant group acting on this does not match the classical one.  
This would parallel the kinematical quantum algebra not matching the entirety of that of the classical observables, 
the physical observables classical and quantum algebras not coinciding as algebras. 
Quantum Configurational Relationalism may require a different group from its classical precursor.

%====================================================================================================================================================================
%===================================================================================================================================================================================
\section{Quantum Assignment of Spacetime Observables (aspect 2$^{\prime}$)}
%===================================================================================================================================================================================
%====================================================================================================================================================================

\n One next determines the zero commutant operators corresponding to the above quantum brackets algebraic structure.  
\be 
\mbox{\bf [} \, \widehat{\bFrF} \mbox{\bf ,} \, \widehat{\Dirac} \, \mbox{\bf ]} \, \Psi \peq 0  
\ee 
is the key equation here.

%===================================================================================================================================================================================
%=========================================================================================================================================================================================
\section{Quantum Foliation Independence (aspect B)}\label{Q-Fol}
%=========================================================================================================================================================================================
%===================================================================================================================================================================================

{\bf Remark 1} A Quantization of GR that retains the nice classical property of Refoliation Invariance would be conceptually sound and widely appealing.   
There is however no known way of guaranteeing this in general at the quantum level.   
If this property is not retained, $\Psi_{\si\sn}$, Kucha\v{r} argued that \cite{K92} starting with the same initial state  
{\it ``on the initial hypersurface and developing it to the final hypersurface along two different routes produces inequality"},
\beq
\Psi_{\sf\si\sn-\sv\si\sa-1} \neq \Psi_{\sf\si\sn-\sv\si\sa-2}  \m \mbox{ (quantum Foliation Dependence criterion) }  .
\label{fol-dep}
\eeq
[This refers to Fig III.6's intermediary hypersurfaces 1 and 2.]
This moreover {\it ``violates what one would expect of a relativistic theory."} 

\m 

\n{\bf Remark 2} On the other hand, association of times with foliations is expected to break down if the spacetime metric 
quantum-mechanically fluctuates as per our concluding discussion.  

\m 

\n{\bf Remark 3} For now, we pose the semiclassical and fully-quantum analogues of Teitelboim's classical resolution by Refoliation Invariance.

%==================================================================================================================================================================================
%====================================================================================================================================================================================						  
\section{Conclusion}
%====================================================================================================================================================================================
%==================================================================================================================================================================================

A Local Resolution of the Problem of Time \cite{K92, I93, Battelle, DeWitt67} was recently put forward \cite{ALett, ABook}.  
This is unprecedented in handling all of the local facets concurrently. 
Further context for this is that these facets have a history of interfering with each other so that solving multiple facets together 
involves much more than the `sum of its parts'.
A Local Resolution of the Problem of Time moreover amounts to establishing A Local Theory of Background Independence \cite{APoT3, ABook}.

\m 

\n \cite{ABook}'s presentation of this local resolution viewed it in three pieces. 
The below Arabic numerals label the Background Independence aspects corresponding to overcoming the Problem of Time facets, 
wheras I-IV and A-B are a grouping elaborated upon in Sec 11.2.  

\m 

\n 0) {\bf Relationalism} of the 0a {\bf Temporal} and 0b {\bf Configurational} kinds. 
Barbour \cite{BB82, B94I} found ways of implementing each of these separately at level of Classical Mechanics and GR actions.
Temporal Relationalism moreover implements earier ideas of Leibniz \cite{L} and Mach \cite{M} via action principles 
whose Mechanics version was first formulated by Jacobi.  
Temporal Relationalism can be resolved via Machian emergent time at both classical and quantum levels, 
providing in particular classical and semiclassical resolutions. 
Configurational Relationalism occurs for Gauge Theory as well; 
the more general idea for handling Configurational Relationalism is the $\lFrg$-act $\lFrg$-all method, of which group averaging is a simple example.  

\m 

\n Both of these Relationalisms are moreover {\bf Constraint Providers}.

\m 

\n 1) Logically, the ensuing constraints next need to be checked for consistency, which constitutes {\bf Constraint Closure}: aspect 1). 
Complications with or failure to attain this closure constitute the {\bf Constraint Closure Problem}, for which novel material was provided in Article III. 
This is addressed along the lines of Dirac's Algorithm \cite{Dirac}. 
This marks the onset of canonical brackets algebras, which then forks into the two following mutually-independent parts. 

\m 

\n 2) {Assignment of Observables} considers what forms zero brackets with the constraints \cite{Dirac49}; 
these observables moreover themselves form brackets algebras. 
Complications with or failure to attain this assignment constritutes the {\it Canonical Problem of Observables}. 

\m 

\n 3) {Constructability of Space from Less Space Structure Assumed}

\m 

\n A) {\bf Running whole families of candidate theories through a Dirac-type Algorithm} -- viewable as algebraic deformation -- 
gives Lie Algebraic Rigidity results, in particular that GR is one of very few cases surviving this consistency check. 
This moreover also provides a means of, firstly, deriving GR-as-Geometrodynamics specific features 
from within ab initio entire families of candidate geometrodynamical theories.  
Secondly, of constructing GR-like spacetime from merely spatial structure alongside the underlying demand of consistency.
This {\bf Spacetime Constructability from Space} is a further Background Independence aspect and Wheelerian route between primalities.   
Starting with less structure than spacetime -- assuming just one or both of spatial structure or discreteness -- 
is particularly motivated by Quantum Theory \cite{Battelle}. 
In such approaches the spacetime concept is to hold at least in suitable limiting regimes.

\m 

\n If this is false, or remains unproven, then we have a {\bf Spacetime Construction Problem}: facet A).

\m 

\n Spacetime constructed, it has aspect 0$^{\prime}$) its own {\bf Relationalism}, 
                                 aspect 1$^{\prime}$) {\bf Closure} -- now of generators: of the spacetime automorphism group -- rather than constraints,   
                                 aspect 2$^{\prime}$) {\bf Assignment of Observables}, for which Article III provides novel material, 
						     and aspect 3$^{\prime}$  {\bf Spacetime from less Spacetime Structure}.  

\m 
							 
\n B) Spacetime can moreover also be foliated and is Refoliation Invariant.
This represents reverse passage: from spacetime primality to space, as opposed to A)'s spatial primality to spacetime.  
It is moreover GR's constraint algebraic structure that implies Refoliation Invariance \cite{Tei73}, 
thus implementing Wheelerian route aspect B): {\bf Foliation Independence}.  

\m 

\n Fig \ref{Evol-Fac} summarizes the progress made in understanding of the Background Independence aspects and Problem of Time facets. 
%
%FFFFFFFFFFFFFFFFFFFFFFFFFFFFFFFFFFFFFFFFFFFFFFFFFFFFFFFFFFFFFFFFFF  F A C E T   N A M E   E N D   P O S I T I O N  FFFFFFFFFFFFFFFFFFFFFFFFFFFFFFFFFFFFFFFFFFFFFFFFFFFFFFFFFFFFFFFFFFFFFF
{            \begin{figure}[!ht]
\centering
\includegraphics[width=1.0\textwidth]{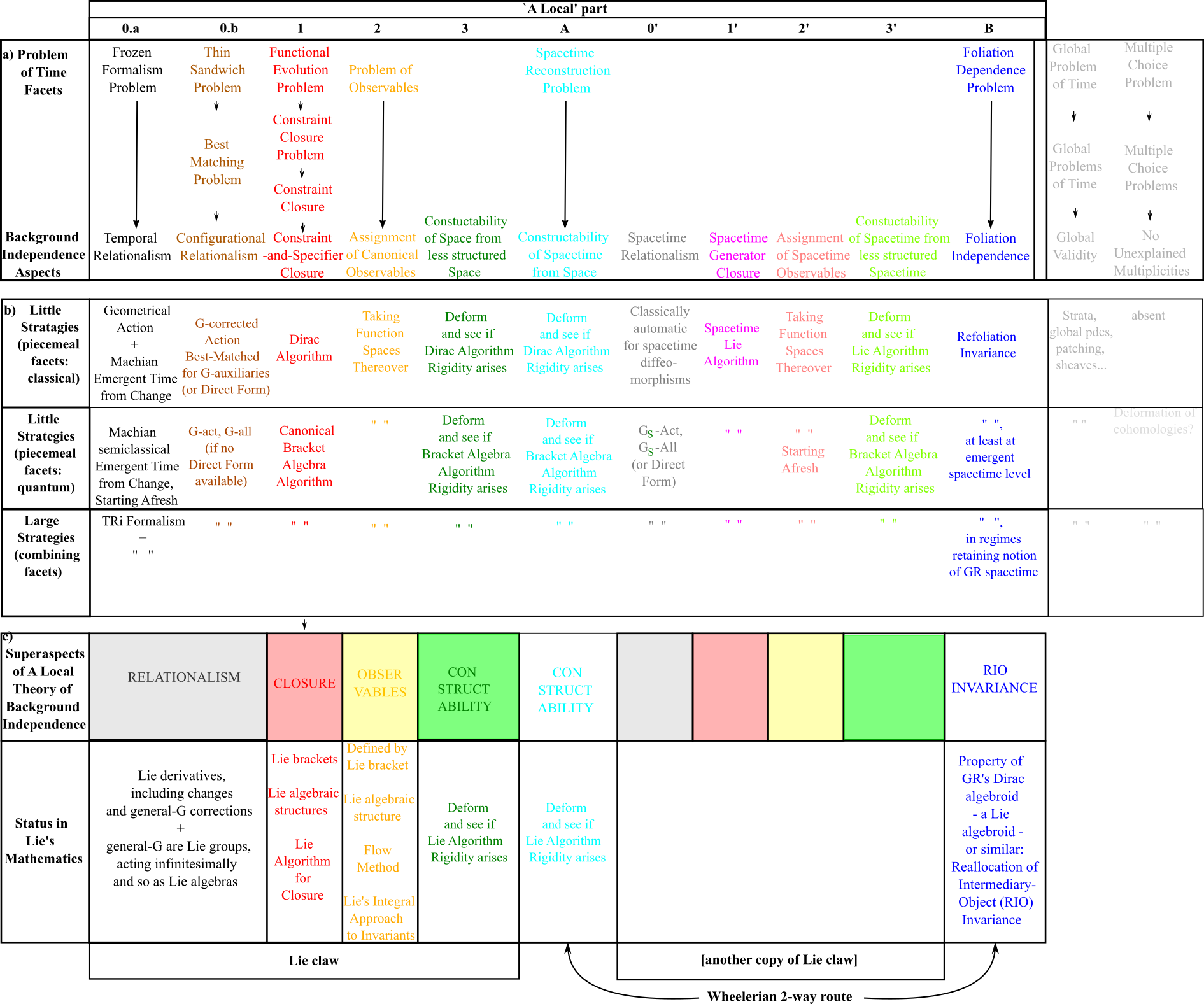}
\caption[Text der im Bilderverzeichnis auftaucht]{\footnotesize{a) Evolution of conceptualization and nomenclature of Problem of Time facets 
into underlying Background Independence aspects over the course of this Series of Articles. 
The first row are Kucha\v{r} and Isham's \cite{K92, I93}.  
The last full row are the underlying Background Independence aspects arrived at in the current Series.
This Figure's colour scheme for the first 11 aspects-and-facets is further used in Articles V to XIII's presentation of interferences therebetween.  
12/13ths of these aspects are moreover already classically present: all bar the issue of physically and conceptually unaccounted-for multiplicities.
See Fig I.2 and Fig II.8 for expansions of column 1 and 2's developments respectively.  

\m 

\n b) The Small Strategies at the Classical and Quantum Levels: for piecemeal aspect incorporation or facet resolution. 
And the Large Strategies for use at whichever level of structure, for whatever theory and robust enough for joint consideration of aspects and facets.

\m 

\n c) The five superaspects: Relationalism, Closure, Observables, Constructability, and RIO Invariance, 
alongside which parts of Lie's Mathematics models each. 
The four pieces of the Lie Claw, and the nature of the Wheelerian 2-way route, are also indicated.
We have thus halved the number of conceptual classes of aspects to subsequently consider: c) is a much simpler state of affairs than a) and b).} }
\label{Evol-Fac}\end{figure}            }
%FFFFFFFFFFFFFFFFFFFFFFFFFFFFFFFFFFFFFFFFFFFFFFFFFFFFFFFFFFFFFFFFFFFFFFFFFFFFFFFFFFFFFFFFFFFFFFFFFFFFFFFFFFFFFFFFFFFFFFFFFFFFFFFFFFFFFFFFFFFFFFFFFFFFFFFFFFFFFFFFFFFFFFFFFFFFFFFFFFFFFFFFF

%====================================================================================================================================================================================
\subsection{Order of approach of facets or aspects}
%====================================================================================================================================================================================
%
%FFFFFFFFFFFFFFFFFFFFFFFFFFFFFFFFFFFFFFFFFFFFFFFFFFFFFFFFFFFFFFFFFFFFFFFFFFFFFFFFFFFFFFFFFFFFFFFFFFFFFFFFFFFFFFFFFFFFFFFFFFFFFFFFFFFFFFFFFFFFFFFFFFFFFFFFFFFFFFFFFFFFFFFFFFFFFFFFFFFFFFFFF
 {            \begin{figure}[!ht]
 \centering
 \includegraphics[width=0.92\textwidth]{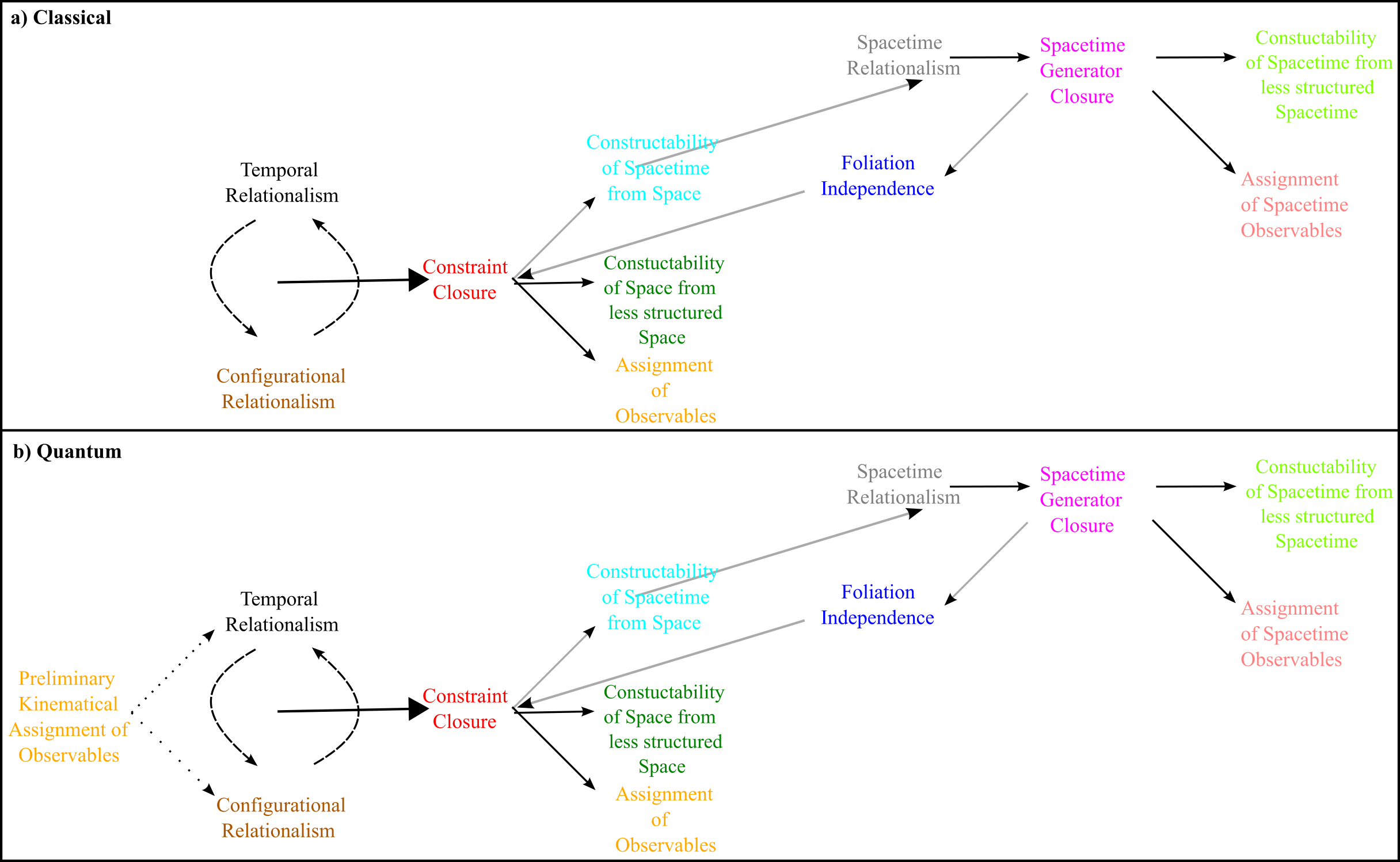} 
 \caption[Text der im Bilderverzeichnis auftaucht]{\footnotesize{a) Classical and b) Quantum order of incorporation of Background Independence aspects, 
i.e. of overcoming corresponding Problem of Time facets.
`Ordering gates' becomes the nonlinear problem of assigning a digraph (`directed graph'), as indicated by the arrows.
b) can be taken to follow a)'s ordering because Lie's Mathematics, and generalizations, transcends to the quantum level.  
These facet orderings are the second major result of the current Series.} } 
 \label{New-Gates} \end{figure}          }
%FFFFFFFFFFFFFFFFFFFFFFFFFFFFFFFFFFFFFFFFFFFFFFFFFFFFFFFFFFFFFFFFFFFFFFFFFFFFFFFFFFFFFFFFFFFFFFFFFFFFFFFFFFFFFFFFFFFFFFFFFFFFFFFFFFFFFFFFFFFFFFFFFFFFFFFFFFFFFFFFFFFFFFFFFFFFFFFFFFFFFFFFF

We now double Fig III.7 to include the corresponding quantum order of addressal. 
The main difference is the prelimnary necessity of kinematical quantization. 
Also note that we strongly argued for emergent Machian time and observables to be found afresh at the quantum level rather than promoted.  

\m 

\n{\sl A Local Resolution of the Problem of Time is to pass consistently through nine facets, ordered as per Fig \ref{New-Gates}}. 
Aspects 3) and 3$^{\prime}$) are, rather, to be viewed as attachment points for less structured possibilities in the Comparative Theory of Background Independence.

%===================================================================================================================================================================
\subsection{Lie's Mathematics \cite{Lie, Serre-Lie, Lee2}}
%===================================================================================================================================================================

\cite{ABook} classified the Background-Independent aspects in various ways. 

\m 

\n C1) By spacetime primality vesus spatial, configurational, dynamical or canonical primality.

\m 

\n C2) By splitting the latter into \cite{ABook} `Barbourville' \cite{BB82, B94I}:    Temporal and Configurational Relationalism, 
                                             and `Diracville' \cite{Dirac, Dirac49}: Constraint Closure and Assignment of Canonical Observables. 

\m

\n We now prefer to view this as a Relationalism--Closure--Observables--Construction {\sl 3-star} (alias {\sl claw graph}) of super-aspects, 
of which Articles III and IV have by now fully argued that spacetime primality carries a separate realization of.  
This is no coincidence because these are {\bf\it Lie 3-stars}, whose existence is {\sl regardless} of these distinctions in primality.  
In particular, 

\m 

\n 0)   All three Relationalisms thus involved -- Configurational, Temporal and Spacetime -- are implemented by {\bf\it Lie derivatives}.  

\m 

\n 1)  Both closures -- Canonical-Constraint and Spacetime-Generator -- involve some type of {\bf\it Lie bracket} 
                                                                         closing consistently as a {\bf\it Lie algebraic structure: generator algebraic structures}. 
[For this purpose, canonical constraints are a subcase of generators.] 																		 
Both are powered by {\bf\it Lie Algorithms}, 
though the canonical subcase of this is for now more widely known, under the name of Dirac Algorithm. 

\m 

\n 2) Both notions of observables are {\bf\it Lie brackets commutants} with the corresponding closed algebraic structures.  
Both Assignments of Observables involve Function Spaces over corresponding spaces -- configuration space and the space of spacetimes -- 
that are moreover themselves {\bf\it Lie algebraic structures: observables algebraic structures}. 

\m 

\n 3) The Constructions of which we speak here are internal, in the sense of space     from less structure of space     assumed 
                                                                          and spacetime from less structure of spacetime assumed. 
All Constructions moreover (including the primality-transcending rather than internal Spacetime from Space) concern {\bf\it putting 
deformed families through the} corresponding {\bf\it Lie Algorithm}, with successes underlied when {\bf Lie Algebraic Rigidity} is encountered.   

\m 

\n Together, each primality's 0 to 3 form the Lie Claw Digraph, with Closure as a central nexus with the other three aspects separately connected to this.  

\m 

\n The finer distinction between our two primalities' 3-stars is as follows. 

\m

\n{\bf Distinction 1)} Separate space (more generally configuration) and time considerations leave us with two Relationalisms.  
This has the knock-on effect of in general needing to cycle between the two until consistency is attained, 
with Closure only then following from both of them (depicted as the thicker `2-multigraph' arrow of Fig 2
).  

\m  

\n{\bf Distinction 2)} Each of these Relationalisms is moreover a constraint provider. 
By which one is led to, explicitly, Constraint Closure (unless specifier equations subsequently appear, 
in which case one would have Constraint-and-Specifier Closure).

\m 

\n{\bf Distinction 3} The canonical approach uses the Poisson (or more generally Dirac) bracket; 
this is more structured than the spacetime Lie bracket because it is also a derivation.   
Also, having a single Relationalism and no appending procedure removes many types of Closure Problem.

\m 

\n{\bf Distinction 4} Canonical observables can then moreover be recast, by Poisson or Dirac brackets' meaning, as 
{\it PDE systems to which the Flow Method applies}, giving an explicit mathematical problem for each theory's canonical observables. 
Spacetime observables can be recast likewise, but now by assigning a differential operator representation for each generator.
(see Articles VIII and X for details).

\m 

\n The remaining two kinds of local aspect realize the Wheelerian \cite{WheelerGRT, Battelle} 2-way passage between our two primalities.

\m 

\n A) Spacetime Construction from Space is passage to spacetime primality from spatial, configurational, dynamical or canonical primality (Article IX).  

\m 

\n B) Foliation Independence is the main part of reverse passage between primalities.  
[Preliminary parts of this reverse passage involve splitting the Relationalism and prescribing foliation kinematics \cite{TRiFol, XII}.]

\m 

\n For GR, this two-way passage is realized as follows.   

\m 

\n Spacetime Construction is afforded by \cite{RWR, AM13, ABook} {\bf Inserting Families of Candidate Theories} into the canonical case's Dirac Algorithm. 
This rests on Lie brackets algebraic structures now exhibiting some {\bf Lie Algebraic Rigidities}, 
returning both details of GR's dynamics and locally-Lorentzian spacetime structure.  
While Dirac Algorithm Rigidities have been known for some years, more general Lie algebra rigidities have recently been discovered in the 
Foundations of Geometry \cite{A-Brackets}, by which this is a truly Lie, rather than just Dirac, phenomenon. 

\m 

\n Foliation Independence is ensured by Refoliation Invariance, which follows \cite{Tei73} from the algebraic form of the {\bf Dirac algebroid} \cite{Dirac51}. 
This is in terms of Poisson brackets -- i.e.\ a type of Lie brackets -- 
while constituting a {\bf Lie algbroid}: a structure of Lie type, albeit exceeding Lie's own work in its complexity.  

\m 

\n{\bf Remark 1} Among the above uses of Lie's Mathematics, it is Lie brackets algebraic structures that see particularly major and central use. 
These are moreover now argued to run not on Dirac's Mathematics but by the more general Lie Mathematics, 
through examples of `Dirac magic' now having been observed for Lie brackets more generally than in Dirac's Poisson brackets setting, 
thereby constituting  `Lie magic'.   

\m 

\n{\bf Remark 2} Overall, Lie's Mathematics is large enough as a setting in which to formulate classical differential-geometric 
and further (semi)Riemannian metric Background Independence, and to solve the associated classical local Problem of Time. 
Indeed, Lie's Mathematics continues to serve in this way in the more general context \cite{AMech, PE16, ABook, A-Killing} of no, or any other, 
geometrical level of structure being assumed.

\m 

\n That Lie's Mathematics is the natural one for local use at Differential-Geometric level of structure, immediately rings true. 
This due to its local character and, within Differential Geometry, universality \cite{Lie}.  

\m 

\n The current Series moreover demonstrates this to be sufficent broad to therein pose and locally resolve the classical-level Problem of Time.
{\sl Lie's Mathematics, as the cornucopia of local Differential Geometry, 
succeeds in providing a full complement of Mathematics for A Local Resolution of the Problem of Time at the classical level}.  

\m 

\n{\bf Remark 3} This clarified, much confusion is lifted, and the classical local Problem of Time becomes far more widely accessible. 
For most Theoretical Physics majors or Grad School freshers know Lie's Mathematics, as do `the continuum half' of their Mathematics counterparts, 
whether pure or applied-geometric, or indeed applying geometry within other parts of STEM \cite{A-Killing}.    
This is in contrast with even Dirac's work on constrained systems unfortunately and undeservedly not being particularly widely known.  
So emphasizing that Lie's Mathematics suffices to conceptually understand a first self-consistent portion of the Problem of Time and Background Independence 
is very useful for this subject. 

\m

\n{\bf Remark 4} We are now in a position to explain the current Series' colouring scheme for facets/aspects/strategies. 
The Lie Claw uses grey for Relationalism, 
                  red-for-emphasis for Closure, 
				  orange for Observables, 
				  and green for Constructability.
These come in two variants: 
       the paler    claw of grey,                                                                      pink, salmon and lime  for spacetime primality, 
versus the brighter claw of black-and-brown (Relationalism's split into Temporal and Configurational), red,  orange and green for canonical primality. 
The Wheelerian 2-way route uses blue: dark blue for Foliation Independence versus cyan for Spacetime Constructability 
(which is part-blue for a Wheelerian route and part-green for a Constructability).  
This colour scheme supercedes \cite{ABook}'s by manifestly realizing the 2 copies of the Lie claw with parallel colouring schemes. 

\m  

\n{\bf Remark 5} Background Independence and the Problem of Time is moreover a very useful subject to know as regards the nature of Physical Law, 
and as regards plenty of interesting open research questions at whichever of the global, quantum or less mathematically structured levels.
						  
\m 

\n{\bf Remark 6} And so the mist has cleared, revealing that   
{\sl the obvious} body of Mathematics serves to frame, and resolve, the Local Problem of Time at the classical level. 
Many conceptual features of this accessible work moreover carry over to the quantum level. 
In this way, `Lie's Mathematics' is not only the cornucopia of Differential Geometry, 
but also the Problem of Time's `alohomora'\cite{Rowling}: a single-word spell for unlocking the gates in question.

%===================================================================================================================================================================
\subsection{The mathematical amphitheatre conditioned by Constrained and Quantum Theories}
%===================================================================================================================================================================

There is the further matter of `mathematical technique' versus the `mathematical amphitheatre' that it is used in.  
The multi-faceted Problem of Time began to be posed in detail in 1967 \cite{Battelle, DeWitt67}, in the context of QM versus GR incompatibility. 
It took until Isham and Kucha\v{r}'s reviews in the early 1990s \cite{K92, I93} to conceptually classify.  

\m 

\n On the one hand, aside from Lie's Mathematics, one needs to know that QM points to 'configurations, 
                                                                                       configuration spaces, 
																					   momenta, 
																					   Poisson brackets, 
																					   phase space, 
																					   Hamiltonians, 
constraints and observables' as useful classical precursors.\footnote{QM was established in the late 1920's. 
%OOOOOOOOOOOOOOOOOOOOOOOOOOOOOOOOOOOOOOOOOOOOOOOOOOOOOOOOOOOOOOOOOOOOOOOOOOOOOOOOOOOOOOOOOOOOOOOOOOOOOOOOOOOOOOOOOOOOOOOOOOOOOOOOOOOOOOOOOOOOOOOOOOOOOOOOOOOOOOOOOOOOOOOOOOOOOOOOOOOO
Furthermore, it took Dirac until the 1950's to understand constraints and observables, 
and it took almost everyone else some decades to catch up with these developments. 
%
% + might eventually balance this and XIV.  
%
Finally, it took even longer for phase space's potential to be unlocked, particularly in a geometrical manner or as regards constraints, let alone for both of these 
to be handled concurrently in a physically satisfactory way.}
%OOOOOOOOOOOOOOOOOOOOOOOOOOOOOOOOOOOOOOOOOOOOOOOOOOOOOOOOOOOOOOOOOOOOOOOOOOOOOOOOOOOOOOOOOOOOOOOOOOOOOOOOOOOOOOOOOOOOOOOOOOOOOOOOOOOOOOOOOOOOOOOOOOOOOOOOOOOOOOOOOOOOOOOOOOOOOOOOOOOO

\m 

\n Remarkably, all of these bar Hamiltonians are already-TRi rather than requiring TRi modification.
The current Article moreover points out that the small modification to classical Hamiltonians is of no subsequent quantum consequence.  
QM then itself involves \cite{RS, KRBook, I84, ABook} kinematical operators, constraint operators, observables operators. 
While the ensuing operator algebras in many ways lie beyond the remit of Lie's own work,  
the quantum commutator involved in these algebras is still mathematically a Lie bracket.

\m 

\n On the other hand, to understand the Problem of Time, one also needs to know some SR (1905), GR (1915), the dynamical structure of GR: constraints, 
                                                                             ADM formulation, 
                                                                             that Hamiltonians and Poisson brackets provide a systematic way of handling constraints 
																		    (from the 1950's and only consolidated in the early 1960's \cite{ADM, Dirac}).
GR also introduced the spacetime versus space--configuration-space--dynamics--phase-space--or--canonical dilemma \cite{ADM, Dirac, Battelle}.
By this, one can moreover consider spacetime versions of brackets and derivative concepts in place of canonical ones.
GR's constraints moreover close as a Lie algebroid rather than a Lie algebra -- the Dirac algebroid -- which is named by analogy with Lie algebras 
rather than lying within the remit of Lie's own work.  

\m 

\n Many of the current subsection's areas are, at least in outline, of common knowledge among Theoretical Physicists, especially those (part-)specializing in GR.  
The most likely exceptions are, firstly, quantum operators, 
for which a background in Functional Analysis is probably more useful than one in Theoretical Physics. 
Secondly, research into the Mathematics of Dirac's algebroid is probably still in its infancy.  

\m 

\n The set of locks to open is in a distinct mathematical amphitheatre to the one in which Lie worked, 
to which, moreover, Lie's Mathematics carries over.  
We moreover also need to substantially reformulate the amphitheatre to succeed in jointly incorporating Background Independence aspects, 
alias jointly resolving Problem of Time facets. 
This is the subject of Articles V to XIII, which show how these reformulations go, as well as that Lie's Mathematics is indeed robust enough to support these too.

%===================================================================================================================================================================
\subsection{The harder parts of A Local Resolution of the Problem of Time}
%===================================================================================================================================================================

From a structural point of view, these are concentrated upon the two-way passage between spacetime 
                                                                                     and `space, configuration, dynamics or canonical' primalities. 
I.e.\, firstly, why and how broadly does Lie Algebraic Rigidity occur as regards Spacetime Construction. 
Secondly, what subset of Lie algebroids encode Refoliation Invariance along the lines that the Dirac algebroid does, 
and how this encodement in general comes about. 
(Is it always a hidden symmetry, and what structure and meaning do such have in general?)  

\m 

\n From a solving point of view, 
explicitly finding spaces of observables remains a formidable task for most constrained, gravitational, Background Independent theories and closed-universe models.
This is moreover to be followed by Expression in Terms of Observables of all of a theory's quantities.  
These two matters remain open questions. 
At the classical level, the latter still lies within the remit of Lie's Mathematics. 
The former may well require current- or future-generation Mathematics to specify a suitable subfield with extra structure within Lie's Mathematics.

%===================================================================================================================================================================
\subsection{Further Frontiers}
%===================================================================================================================================================================

\n Recollect from Article I's Introduction that the final two facets of the Problem of Time are as follows. 

\m 

\n Aspect 11) is {\bf Global Validity}, difficulties with which are termed the {\bf Global Problems of Time} 
(facet 11: see \cite{K92, I93}, Epilogues II.B and III.B of \cite{ABook} and \cite{A-Killing, A-Cpct, A-CBI, Higher-Lie}).  
Global Problems occur with all facets, by which outlining these is better to detail this after facet interferences have been covered, 
so see Article XIV's Conclusion.   

\m 

\n Aspect 12) is {\bf Unexplained Multiplicities}.  
difficulties with which are termed the {\bf Multiple Choice Problems of Time} (facet 12: see \cite{K92, I93, Gotay00} and Epilogue II.C of \cite{ABook}, 
the main part of which would appear to be the quantum-level Groenewold--van Hove phenomenon).
Multiple-choice Problems are particularly severe in, 
specifically, the passage from classical to quantum, and so are avoided in purely classical treatments.
These can already be envisaged for Kinematical Quantization but affect some other facets as well (in particlar the Problems of Observables).  

\m 

\n{\bf Background Independence and the Problem of Time to further levels of mathematical stucture} 
e.g.\ topological manifolds, topological spaces, or sets, is a further research direction (Epilogue II.C of \cite{ABook}). 

\m 

\n{\bf Universal aspects and universal strategies} Some Problem of Time strategies are {\it universal}, 
in the sense that they exist regardless of what the underlying theory is, at least for a large number of steps.  
For instance, universality applies to Emergent Machian Times Approaches, Timeless Approaches and Histories Approaches. 
A further consequence of universality is that Parts II and III are not only of interest for the specific theories used there as examples, 
but also of much wider interest throughout Gravitational Theory.  
While the differential-geometric level of structre is implied by using Lie's Mathematics, it does admit a discrete-differences \cite{DH13} version. 
We also term study of Background Independence at other levels of mathematical structure 
{\bf Comparative Theory of Background Independence} \cite{A-CBI}.

\m 

\n The Global, Quantum and Comparative research directions can moreover be considered in any combination. 
The Multiple Choice Problem also impinges upon each quantum treatment. 

\m

\n We consider it prudent to handle classical-level local joint resolution of Problem of Time facets within the remit of Lie's Mathematics 
in Articles V to XIII prior to discussing Global, Quantum and Comparative generalizations, as these require far more than just Lie's Mathematics. 
As such, we only return to global and less structured matters in Article XIV, and not at all within the current Series to quantum matters.
Subsequent series of Articles on a local resolution of the Quantum Problem of Time and on the classical Global Problem of Time are moreover imminent.  
TRiCQT, TRiPIQT and semiclassical resolution are already in print -- see Part III of \cite{ABook} -- 
as are some Global and Comparative quantum matters: see Epilogues III.B-C of \cite{ABook}.

%=====================================================BIBLIOGRAPHY==========================================================================================================================

\end{document}